\documentclass[twocolumn, tighten, times]{aastex7}

\usepackage{amsmath}

\begin{document}

\title{Calibration of the Next Generation Palomar Spectrograph as a Stellar Speedometer}

\author[orcid=0000-0002-1386-0603,gname=Pranav,sname=Nagarajan]{Pranav Nagarajan}
\affiliation{Department of Astronomy, California Institute of Technology, 1200 E. California Blvd., Pasadena, CA 91125, USA}
\email[show]{pnagaraj@caltech.edu}  

\author[orcid=0000-0002-6871-1752,gname=Kareem,sname=El-Badry]{Kareem El-Badry}
\affiliation{Department of Astronomy, California Institute of Technology, 1200 E. California Blvd., Pasadena, CA 91125, USA}
\email{kelbadry@caltech.edu}  

\begin{abstract}

We calibrate the Next Generation Palomar Spectrograph (NGPS) for precision stellar radial-velocity (RV) measurements using repeated observations of 24 Gaia-ESO RV standards spanning $G=11$--$18$. Because NGPS is mounted at the Hale telescope's Cassegrain focus, changes in instrumental flexure produce wavelength-solution shifts of tens of km~s$^{-1}$, with substantial time and wavelength dependence. We develop a pipeline that measures these shifts directly from each science exposure, using telluric absorption features in the R and I channels and the [O~I] 5577~\AA\ sky-emission line in the G channel. This approach eliminates the need for frequent on-sky arc calibrations. The resulting wavelength-dependent flexure corrections reduce the median absolute discrepancy relative to Gaia-ESO RVs from tens of km~s$^{-1}$ to 1.6~km~s$^{-1}$. Telluric absorption provides substantially more reliable flexure calibration than sky emission lines, which do not track point-source illumination of the slit. The I channel, which contains both many telluric lines and strong photospheric lines across a range of spectral types, generally yields the most stable RVs. The achieved precision remains above the photon-noise limit, indicating that NGPS RV performance is currently limited by calibration and modeling systematics. These results demonstrate that NGPS is well-suited for accurate and precise stellar RV measurements across a wide range of astrophysical studies.
\end{abstract}



\section{Introduction}
\label{sec:intro}

Radial velocity (RV) measurements via Doppler spectroscopy are important for many areas of astronomy: they are used to detect exoplanets, characterize binary stars, probe the evolution of explosive transients, and measure the expansion of the Universe. Spectroscopic follow-up represents the gold standard in confirming the detection of candidate black holes or neutron stars in binaries discovered via other methods (i.e.\ X-rays, photometry, or astrometry; e.g., \citealt{bolton_identification_1972, webster_murdin_1972, mcclintock_remillard_1986, casares_jonker_2014, nagarajan_spectroscopic_2023, green_upper_2025, el-badry_red_2023, el-badry_population_2024, nagarajan_low_2025, lam_search_2026, simon_rv_2026}). Since such binaries are rare and false positives abundant, campaigns to discover these systems require an RV ``engine,'' or a spectrograph that can efficiently and accurately derive RVs for stars over a wide range of apparent magnitudes and spectral types. Furthermore, several upcoming datasets, such as \textit{Gaia} DR4, Rubin LSST, and Roman GBTDS are expected to discover large numbers of interesting binaries, and multi-epoch RV follow-up will be essential to characterize them. 

In this work, we investigate the capability of the Next Generation Palomar Spectrograph (NGPS, \citealt{jiang_ngps_2018}, Fremling et al.\ in prep.), a medium-resolution multi-band instrument on the Hale 200-inch telescope at Palomar Observatory, to perform precise RV follow-up of relatively faint targets of interest. In 2025, NGPS replaced the Double Spectrograph (DBSP, \citealt{Oke_1982}) as Palomar's workhorse spectrograph. NGPS features an adjustable slit width and achieves a spectral resolution of $R \approx 4000$ at a slit width of $0.5''$. With a high signal-to-noise spectrum, the photon-limited RV precision of such an instrument is predicted to be $\lesssim 1\,\rm km\,s^{-1}$ \citep[][]{bouchy_pepe_queloz_2001}, suitable for characterizing the orbits of most binaries being discovered with {\it Gaia}. Unfortunately, this precision is difficult to achieve in practice because the instrument's wavelength solution changes significantly on timescales of minutes to years. These changes are a result of a variety of factors, including short-timescale mechanical flexure in the instrument, changes in the ambient temperature, and miscentering of the star in the slit.

A common but suboptimal practice to account for flexure is to spend precious telescope time taking an arc on sky immediately before and/or after each science target.\footnote{This approach was even recommended in the DBSP cookbook itself --- see Section 7 of \url{https://sites.astro.caltech.edu/palomar/observer/200inchResources/dbspcookbook.html}.} Another common practice is to compare the observed and expected wavelengths of sky emission lines imprinted on the spectra of astronomical objects and use these to ``correct'' the wavelength solution \citep[e.g.,][]{wilson_apogee_2019, perley_lpipe_2019}. Unfortunately, the pixel-to-wavelength mapping for starlight is not necessarily the same as the mapping for arcs or sky lines, even if the star is perfectly centered in the slit. This is because a point source like a star illuminates the slit differently than does a lamp; the latter fills the slit, while the former does not. For this same reason, corrections to the wavelength solution using sky emission lines usually fall well short of the theoretically achievable precision as well \citep[e.g.,][]{griest_hires_2010}.   

Fortunately, nature provides a solution in the form of telluric absorption lines superposed on the spectra of astronomical sources. These lines are ideal for calibrating shifts in the wavelength solution during the course of a night because they are imprinted directly on the source spectrum \citep[e.g.,][]{griffin_possibility_1973, simon_geha_2007, geha_deimos_2026}. This means that their positions accurately track the mean integrated slit function. We have previously used the O$_2$ A-band to stabilize RVs measured with DBSP \citep{nagarajan_spectroscopic_2023}. In this work, we develop a wavelength-dependent flexure correction pipeline for NGPS, calibrating it as a stellar speedometer based on observations of 24 RV standard stars of apparent magnitude $G = 11$--$18$ from the Gaia-ESO spectroscopic survey \citep{randich_eso_2022}. We probe NGPS's RV stability using repeated observations separated by several months, and additionally test the effect of slit width on the instrument's RV precision.

The remainder of this paper is organized as follows. In Section \ref{sec:data}, we describe our campaign to obtain NGPS spectra of Gaia-ESO RV standard stars spanning a range of magnitudes. In Section~\ref{sec:methods}, we describe the wavelength-dependent flexure correction method implemented in our RV measurement pipeline. In Section~\ref{sec:results}, we apply this pipeline to derive robust RVs in the G, R, and I channels for all of our targets. In Section \ref{sec:discussion}, we discuss the RV performance and stability of NGPS and provide practical recommendations for effective use of our pipeline. Finally, in Section \ref{sec:conclusion}, we explore directions for future work.

\section{Data}
\label{sec:data}

\subsection{Sample Selection}
\label{sec:sample_selection}

We selected 24 candidate RV standard stars spanning \textit{Gaia} apparent magnitudes $G = 11$--$18$ from the final public data release of the Gaia-ESO spectroscopic survey \citep{randich_eso_2022}. Specifically, we identified stars with Gaia-ESO RV uncertainties $< 0.3$ km s$^{-1}$. We additionally restricted our sample to stars with effective temperatures between $4000$ K and $6000$ K and metallicities [Fe/H] $>-0.5$. Most of our targets were main-sequence dwarfs, but we also observed a few evolved stars (i.e., subgiants or red giants). We favored relatively sparse fields, and only selected targets observable from Palomar during our campaign.

As an additional check against multiplicity, we compared the Gaia-ESO RVs with \textit{Gaia} DR3 catalog RVs \citep{katz_gaia_2023} where available. For targets with both Gaia-ESO and \textit{Gaia} DR3 RVs, we verified that they are statistically consistent given their combined uncertainties. We also confirmed that none of these targets has a published solution in the DR3 non-single star catalogs \citep{gaia_collaboration_gaia_2023-1}. These checks reveal no direct evidence for binarity or substantial RV variability, though we cannot exclude unresolved or long-period companions. Since we cannot rule out the possibility that a few of these objects are binaries with detectable RV variability, the empirical RV uncertainties we infer are upper limits. Henceforth, we refer to all targets as ``RV standards.'' We provide a log of our observed targets in Table \ref{tab:all_rvs}. 

\subsection{Observing Campaign}
\label{sec:campaign}

We obtained NGPS spectra of 17 of these Gaia-ESO RV standards in the two available channels (R and I) in commissioning mode on UT date July 21, 2025. We obtained NGPS spectra of 14 of these sources and 7 additional standards in all four channels (U, G, R, and I) on UT date May 15, 2026. The U, G, R, and I channels have nominal bandpasses of 3050--4430, 4250--5960, 5620--7950, and 7530--10400~\AA, respectively.\footnote{\label{fn:ngps-specs}\url{https://caltechopticalobservatories.github.io/NGPS/technical-specifications.html}} On the first night, we used two slit widths, $0.5''$ and $0.7''$, to test the effect of slit width on RV precision. On the second night, we used slit widths of $0.5''$ and $0.36''$. The latter approaches the Nyquist sampling criterion, or the limit in which a resolution element is sampled by two detector pixels, in the I channel. We used exposure times between $300$--$1800$~s based on apparent magnitude, with longer exposure times for fainter targets. In total, we obtained 62 observations: 33 two-channel observations in 2025 and 29 four-channel observations in 2026.


We estimated the instrumental resolution from the full-width half-maxima of strong, isolated sky emission lines. At $8920$\,\AA, we obtained an average spectral resolution of $R \approx 3000$ with the $0.7''$ slit, $R \approx 4000$ with the $0.5''$ slit, and $R \approx 4800$ with the $0.36''$ slit. At $6300$\,\AA, the average resolution was lower: $R \approx 2500$ with the $0.7''$ slit, $R \approx 3000$ with the $0.5''$ slit, and $R \approx 3700$ with the $0.36''$ slit, respectively. These values are broadly consistent with instrument specifications.\textsuperscript{\ref{fn:ngps-specs}}

We measured the signal-to-noise ratio (SNR) of each spectrum using DER\_SNR \citep{stoehr_dersnr_2008}, a model-independent algorithm that estimates the signal from the median flux and the noise from the scaled median absolute third-order flux difference (computed two pixels apart). Across both observing nights and all slit widths, the median SNR decreased from approximately 33, 53, and 68 for targets with $11\leq G<13$ (for which we typically used 300 s exposures) to approximately 17, 28, and 27 for targets with $17\leq G\leq18$ (for which we used longer 1200--1800 s exposures) in the G, R, and I channels, respectively. We found that SNR increased modestly with slit width: for paired observations of the same targets, the average SNR across available channels was $\approx 12$\% higher with the $0.70\arcsec$ slit than the $0.50\arcsec$ slit in 2025 and $\approx 26$\% higher with the $0.50\arcsec$ slit than the $0.36\arcsec$ slit in 2026.

\begin{deluxetable*}{cccccccccc}
\tablecaption{Log of NGPS observations of Gaia-ESO RV standards. Right ascension and declination can be read directly from the target names. For observations using multiple slit widths, the listed exposure time applies to each slit width. NGPS was used in 2-channel commissioning mode (R and I) on July 21, 2025 and in 4-channel mode (U, G, R, and I) on May 15, 2026.
\label{tab:all_rvs}}
\tablehead{
\colhead{Target} & \colhead{$G$} & \colhead{$T_{\rm eff}$} &
\colhead{$\log g$} & \colhead{[Fe/H]} & \colhead{Gaia-ESO RV} &
\multicolumn{2}{c}{UT 2025 Jul 21} &
\multicolumn{2}{c}{UT 2026 May 15} \\
\colhead{} & \colhead{} & \colhead{} & \colhead{} & \colhead{} &
\colhead{} & \colhead{Slit widths} & \colhead{$t_{\rm exp}$} &
\colhead{Slit widths} & \colhead{$t_{\rm exp}$} \\
\colhead{} & \colhead{mag} & \colhead{K} & \colhead{cgs} &
\colhead{dex} & \colhead{km s$^{-1}$} &
\colhead{arcsec} & \colhead{s} & \colhead{arcsec} & \colhead{s}
}
\startdata
\cutinhead{Observed on both nights}
GES J14595199-0453010 & 15.41 & 5541 & 4.50 & $-0.22$ & $-56.83\pm0.14$ & 0.50, 0.70 & 900 & 0.50 & 900 \\
GES J15015015-0953433 & 14.88 & 5280 & 4.53 & $-0.15$ & $22.29\pm0.11$ & 0.50, 0.70 & 900 & 0.50 & 900 \\
GES J15291525-0456248 & 14.22 & 5182 & 4.55 & $-0.29$ & $-16.45\pm0.11$ & 0.50, 0.70 & 600 & 0.50 & 600 \\
GES J15492188-0731041 & 14.61 & 4692 & 4.62 & $-0.28$ & $32.76\pm0.11$ & 0.50, 0.70 & 900 & 0.50 & 900 \\
GES J17443177+0518564 & 12.98 & 4660 & 4.69 & $0.04$ & $11.05\pm0.25$ & 0.50, 0.70 & 300 & 0.36, 0.50 & 300 \\
GES J17452610+0515305 & 12.53 & 5986 & 4.16 & $-0.06$ & $-60.32\pm0.25$ & 0.50, 0.70 & 300 & 0.36, 0.50 & 300 \\
GES J17461465+0542155 & 12.12 & 5396 & 4.53 & $0.09$ & $-23.42\pm0.24$ & 0.50, 0.70 & 300 & 0.36, 0.50 & 300 \\
GES J17470337+0531556 & 13.47 & 5150 & 4.47 & $-0.08$ & $-27.10\pm0.24$ & 0.50, 0.70 & 600 & 0.36, 0.50 & 600 \\
GES J17472494+0533106 & 13.02 & 5635 & 4.19 & $-0.01$ & $-95.27\pm0.25$ & 0.50, 0.70 & 300 & 0.36, 0.50 & 300 \\
GES J17474442+0530005 & 11.97 & 5723 & 4.17 & $0.11$ & $-24.82\pm0.24$ & 0.50, 0.70 & 300 & 0.36, 0.50 & 300 \\
GES J18264217+0627514 & 11.65 & 5556 & 4.26 & $-0.11$ & $55.18\pm0.10$ & 0.50, 0.70 & 300 & 0.36, 0.50 & 300 \\
GES J18274122+0630205 & 11.74 & 4714 & 2.63 & $0.44$ & $-18.16\pm0.10$ & 0.50, 0.70 & 300 & 0.50 & 300 \\
GES J19253812+0016526 & 13.75 & 4086 & 1.74 & $-0.25$ & $6.62\pm0.11$ & 0.50, 0.70 & 600 & 0.36, 0.50 & 600 \\
GES J21092991-0156380 & 16.03 & 4824 & 4.54 & $-0.49$ & $-43.79\pm0.15$ & 0.50, 0.70 & 900 & 0.50 & 900 \\
\cutinhead{Observed only on UT 2025 July 21}
GES J22092917-0556451 & 17.51 & 4375 & 4.59 & $-0.38$ & $-11.49\pm0.23$ & 0.50 & 1200 & \nodata & \nodata \\
GES J22495053-0453236 & 16.42 & 5488 & 4.56 & $-0.30$ & $-70.67\pm0.19$ & 0.50, 0.70 & 1200 & \nodata & \nodata \\
GES J23401456-0056196 & 17.04 & 5358 & 4.27 & $-0.47$ & $70.72\pm0.26$ & 0.50, 0.70 & 1200 & \nodata & \nodata \\
\cutinhead{Observed only on UT 2026 May 15}
GES J13011804-0602308 & 13.88 & 5330 & 4.58 & $-0.20$ & $-36.75\pm0.10$ & \nodata & \nodata & 0.50 & 600 \\
GES J13200697-0849187 & 16.55 & 4794 & 4.53 & $-0.43$ & $74.42\pm0.19$ & \nodata & \nodata & 0.50 & 1200 \\
GES J14271865-0901055 & 17.46 & 4831 & 4.44 & $-0.24$ & $-42.31\pm0.27$ & \nodata & \nodata & 0.50 & 1800 \\
GES J15392822-0503425 & 16.90 & 5926 & 4.58 & $-0.43$ & $-30.00\pm0.11$ & \nodata & \nodata & 0.50 & 1200 \\
GES J15393205-0743239 & 15.18 & 4817 & 3.13 & $-0.35$ & $-80.23\pm0.11$ & \nodata & \nodata & 0.50 & 900 \\
GES J15400600-0740002 & 17.84 & 5261 & 4.50 & $-0.28$ & $-118.41\pm0.29$ & \nodata & \nodata & 0.50 & 1800 \\
GES J15501564-0731478 & 17.14 & 5166 & 4.44 & $-0.41$ & $-63.54\pm0.24$ & \nodata & \nodata & 0.50 & 1200
\enddata
\end{deluxetable*}

\subsection{Spectral Extraction}
\label{sec:extraction}

\begin{figure*}
    \centering
    \includegraphics[width=\textwidth]{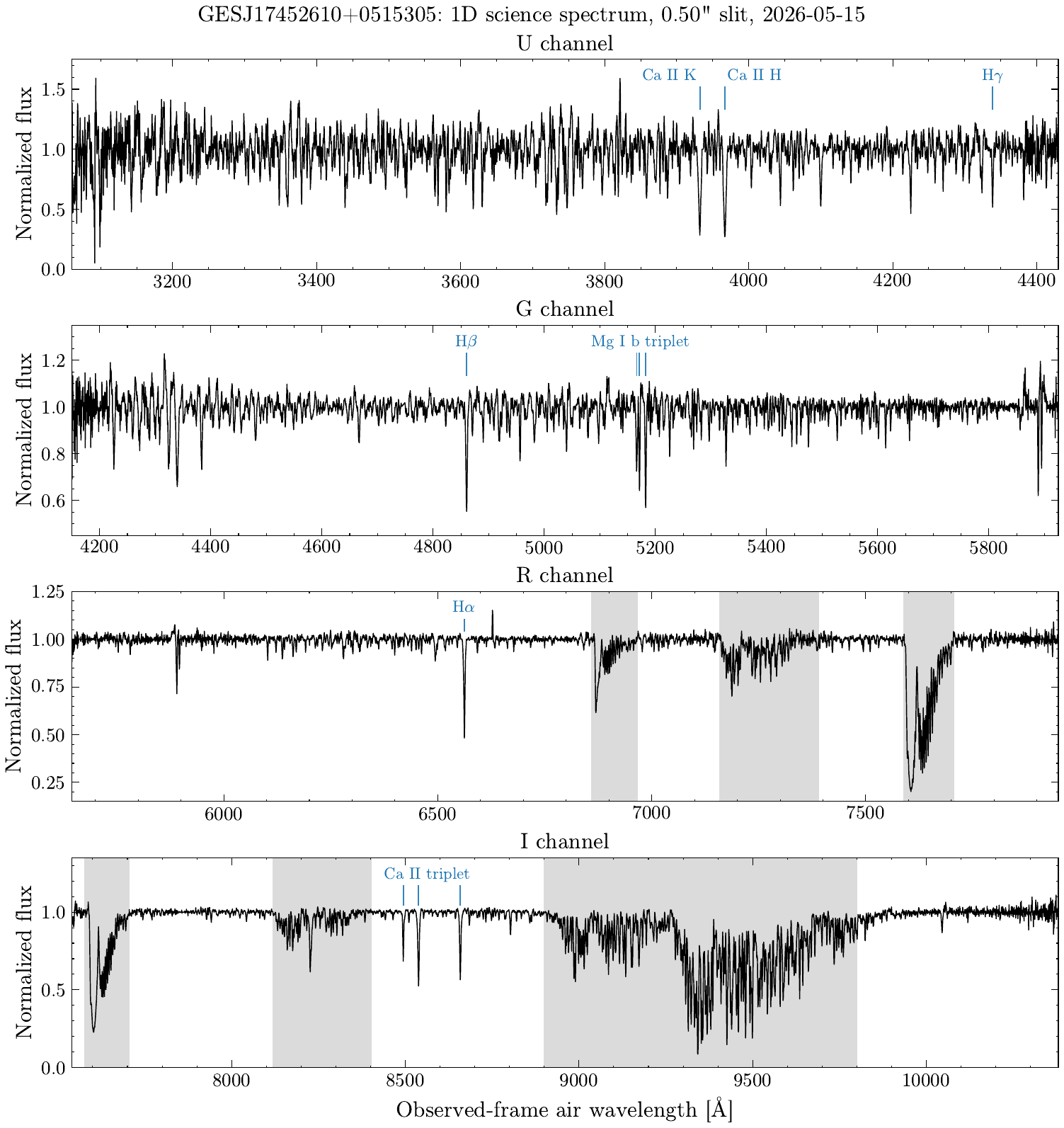}
    \caption{Normalized spectra in the U, G, R, and I channels of an example RV standard star observed with the $0.5''$ slit. Regions with strong telluric absorption features, due to water vapor and oxygen, are shaded in gray. Prominent stellar absorption lines are labeled with blue markers.}
    \label{fig:full_spectra}
\end{figure*}

\begin{figure*}
    \centering
    \includegraphics[width=\textwidth]{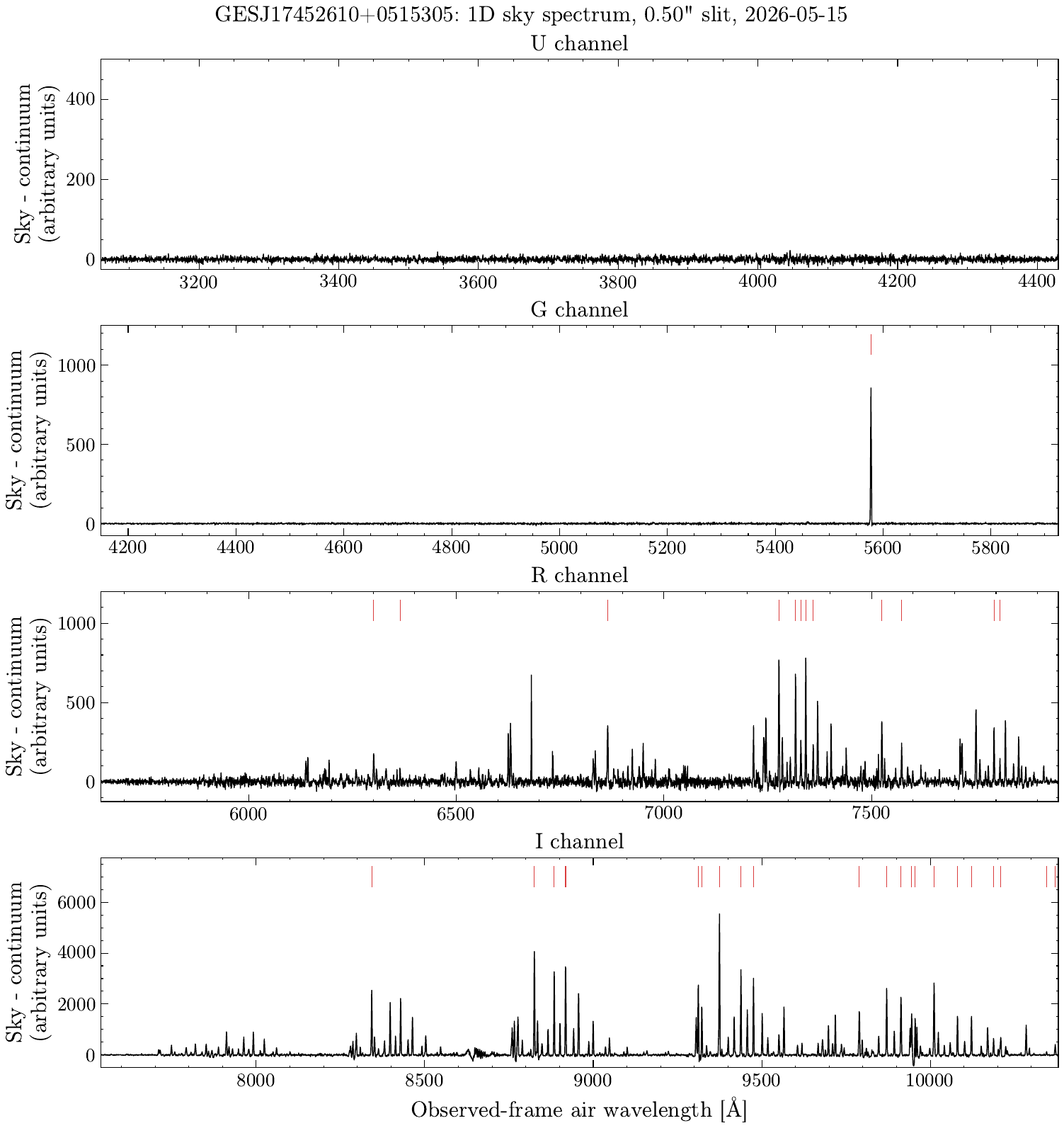}
    \caption{Example sky spectra in the U, G, R, and I channels, extracted from an NGPS observation of the standard star in Figure~\ref{fig:full_spectra}. Telluric emission features become stronger at redder wavelengths. Sky lines used as anchors for flexure measurements are indicated with red markers. Not all emission lines are marked, because several lines in the \citet{hanuschik_emission_2003} catalog are filtered out by strength and separation cuts.}
    \label{fig:sky_lines}
\end{figure*}

We reduced the 2D spectra with the NGPS QuickLook Data Reduction Pipeline (DRP),\footnote{\url{https://caltechopticalobservatories.github.io/NGPS/users-manual/data-reduction-and-telemetry.html}} which performs bias subtraction and flat-field correction, derives the wavelength solution, and estimates and subtracts a sky emission model. We additionally cleaned the reduced 2D spectra of cosmic rays using the L.A.Cosmic algorithm \citep{la_cosmic_2001}, with flagged pixels being replaced by a local masked median.

The focal plane of NGPS is sub-divided into three image slices, with the side slices designed to collect light that would otherwise be lost at the edges of the slit. Since the slices exhibit small relative offsets between their wavelength solutions (see Appendix~\ref{sec:appendix}), we considered only the central slice, which captures most of the signal for a target centered on the slit. We identified the location of the science trace in this slice as the maximum of the spatial profile, constructed from the median flux at each spatial pixel. We extracted 1D spectra of each target in each channel by performing a boxcar extraction around this location with an aperture of width 5 pixels. The QuickLook DRP does not currently include a 2D noise model, precluding an optimal Horne extraction \citep[][]{horne_optimal_1986}. Instead, we initially estimated the relative flux uncertainty scale from the inverse of the measured SNR. 

We also extracted 1D sky spectra from each 2D sky model, adopting the same boxcar aperture and central spatial pixel as the science spectra. We found that the location of the centroid of a given sky emission line varied linearly over the 5-pixel boxcar aperture, so taking the sum over the aperture caused any small local spatial tilt to cancel out. We investigate the spatial variation of sky emission lines in more detail in Appendix~\ref{sec:appendix}.

We show an example of a pseudo-continuum normalized NGPS spectrum from our program, taken in 4-channel mode with the $0.5''$ slit, in Figure~\ref{fig:full_spectra}. We shade regions corresponding to strong telluric absorption features (e.g., oxygen or water vapor bands) in gray. Prominent absorption lines commonly used for RV determination include the Ca II triplet in the I band, H$\alpha$ in the R band, H$\beta$ and the Mg Ib triplet in the G band, and H$\gamma$ and the Ca II H and K lines in the U band. While we use information from the full spectral range in our RV pipeline, we label these stellar lines in Figure~\ref{fig:full_spectra} to emphasize important spectral regions in each channel.

We show examples of extracted 1D sky spectra in the U, G, R, and I channels (for the same standard star as in Figure~\ref{fig:full_spectra}) in Figure~\ref{fig:sky_lines}. Telluric emission features become stronger at redder wavelengths, with the R and I channels showing several strong sky lines. The sky spectrum in the G channel is dominated by the strong 5577\,\AA~[O I] emission line, while the U channel does not show any useful emission features. Emission lines used by our pipeline to measure flexure corrections (i.e., as a fallback option) are identified with red markers (see Section~\ref{sec:sky_emission}).

\section{Methods}
\label{sec:methods}

We measured robust RVs from reduced NGPS spectra by solving two linked problems: (1) measurement of the (wavelength-dependent) flexure correction and (2) measurement of the stellar RV after applying that correction. In the R or I channels, telluric absorption features were used as the default source for measuring flexure, with sky emission lines treated as a fallback option. In the G channel, flexure was measured from the [O I] 5577\,\AA~sky emission line, as no strong telluric absorption features exist. We describe our flexure correction and RV determination procedures in more detail in Sections~\ref{sec:flexure_corrections} and \ref{sec:radial_velocities}, respectively.

\subsection{Flexure Corrections}
\label{sec:flexure_corrections}

\subsubsection{Telluric Absorption Features}
\label{sec:absorption}

Because NGPS is mounted at the Cassegrain focus of the Hale telescope, changes in telescope orientation can produce mechanical flexure and corresponding shifts in the wavelength solution. To compensate for these shifts, we used telluric absorption lines with known rest-frame wavelengths to determine a wavelength-dependent flexure correction at each epoch. 

Specifically, for each observation, we followed the approach of \citet{nagarajan_spectroscopic_2023} to generate a grid of high-resolution telluric model spectra at different airmasses (with spacing $\Delta X = 0.05$) from the HITRAN2020 molecular spectroscopic database \citep{gordon_2022}. We then selected the telluric model spectrum with the closest airmass to the observation. We smoothed this template to the spectral resolution of the observation assuming a Gaussian line-spread function. In doing so, we adopted a constant resolving power across each channel, set to the median value measured from strong, isolated sky emission lines in that exposure (see Section~\ref{sec:campaign}). Next, we identified windows with ``strong'' telluric absorption features as regions where the shifted telluric model predicted absorption depth $\geq 8\%$ (atmospheric transmission $\leq 92\%$).\footnote{Admitting shallower features degraded the performance of the pipeline, while requiring deeper features produced no further improvement.} To normalize the observed spectrum in such regions, we divided by a local pseudo-continuum estimated with a 101\,\AA~median filter. In regions overlapping the O$_2$ A-band, a substantial fraction of the window is occupied by telluric absorption, biasing the rolling median estimate. In these regions, we instead estimated the continuum from the 98th-percentile flux in 30\,\AA-wide bins using a smoothed, shape-preserving piecewise cubic interpolation. We assumed that the continuum-normalized observed flux $F(\lambda)$ in a window of central wavelength $\lambda_{\rm win}$ can be described by a local multiplicative model:

\begin{equation}
\begin{aligned}
F(\lambda) \simeq C(\lambda) & \left\{1+a_{\star}\left[S\!\left(\lambda;v_{\star, 0}-v_{\rm bary}+v_{\rm flex}(\lambda_{\rm win})\right)-1\right]\right\} \\ & \left\{1+a_T\left[T\!\left(\lambda;v_{\rm flex}(\lambda_{\rm win})\right)-1\right]\right\},
\end{aligned}
\end{equation}

\noindent where $S$ is the smoothed stellar template (see Section~\ref{sec:radial_velocities}), $T$ is the smoothed telluric template, $v_{\star, 0}$ is a (fixed) preliminary estimate of the stellar RV, $v_{\rm bary}$ is the barycentric correction, and $v_{\rm flex}(\lambda_{\rm win})$ is the local flexure correction evaluated at the central wavelength. In addition, $C(\lambda)$ is a linear continuum correction that accounts for any remaining offset or slope after normalization, and $a_\star$ and $a_T$ are nuisance parameters that scale the apparent depths of the stellar and telluric features, respectively. In our convention, the flexure correction is subtracted from the observed barycentric RV to obtain the true stellar RV.

To determine $v_{\star, 0}$, we derived a preliminary flexure curve based on strong sky emission lines (see Section~\ref{sec:sky_emission}), and measured a preliminary stellar RV using those flexure measurements (see Section~\ref{sec:radial_velocities}). Then, we held $v_{\star,0}$ fixed and fitted the local flexure correction (along with $C(\lambda)$, $a_{\star}$, and $a_T$) in each window. Because these windows were selected for strong telluric absorption and often contained few useful stellar features, they did not independently constrain both velocities well: a change in the stellar RV could instead be absorbed into the fitted flexure correction. We therefore derived the flexure curve first and subsequently measured the final stellar RV from regions dominated by stellar absorption features (see Section~\ref{sec:radial_velocities}).

We derived the local flexure correction anchored at $\lambda_{\rm win}$ by minimizing the squared residuals between the normalized observed spectrum and the multiplicative model. We fitted the windows in order of increasing wavelength, and initialized $v_{\rm flex}$ by comparing the observed spectrum with the telluric template in a region with strong telluric absorption --- by default, the O$_2$ B band in the R channel and the O$_2$ A band in the I channel. Each converged fit supplied the initialization for $v_{\rm flex}$ in the next window. We retained a flexure correction if the flexure uncertainty was $< 35$ km s$^{-1}$. Additionally, we rejected flexure measurements that were $4\sigma$ outliers relative to the median flexure correction.

\begin{figure*}
    \centering
    \includegraphics[width=\textwidth]{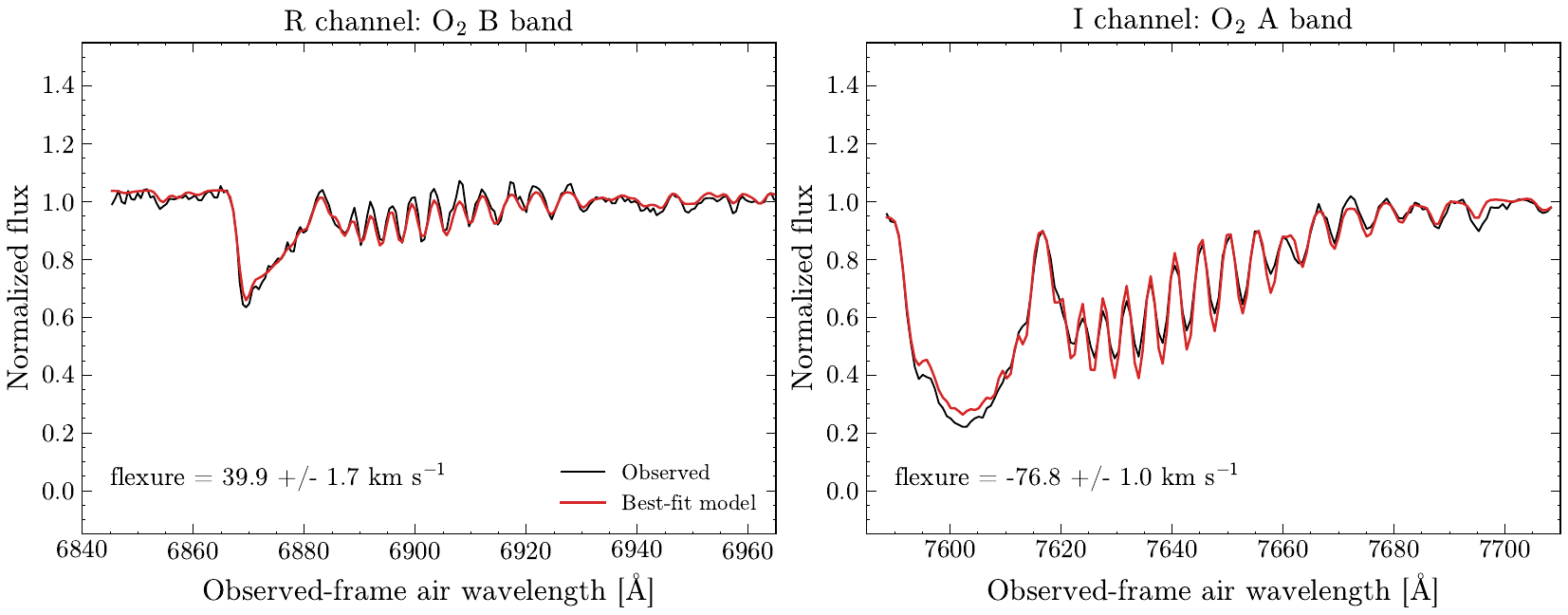}
    \caption{Examples of measurements of local flexure corrections in the R and I channels based on telluric absorption features, focusing on the O$_2$ B-band (left) and A-band (right), respectively. The best-fit stellar + telluric multiplicative model, plotted in red, is compared against the observed spectrum, plotted in black. The measured flexure is $39.9 \pm 1.7$ km s$^{-1}$ around the O$_2$ B-band and $-76.8 \pm 1.0$ km s$^{-1}$ around the O$_2$ A-band, revealing that the best-fit flexure correction varies strongly from channel to channel.}
    \label{fig:flexure}
\end{figure*}

We show examples of local flexure correction measurements in Figure~\ref{fig:flexure}. We compare the normalized observed spectrum in black against the best-fit multiplicative model in red. The flexure correction derived in the window centered on the O$_2$ B band in the R channel is $39.9 \pm 1.7$ km s$^{-1}$, while the flexure correction derived in the window centered on the O$_2$ A band in the I channel is $-76.8 \pm 1.0$ km s$^{-1}$. Clearly, the best-fit flexure correction differs strongly between channels.

To create a flexure curve in the R or I channels, we linearly interpolated between anchor regions where strong telluric features were present. Outside of the supported wavelength range, the flexure was clamped to the value at the nearest edge. In the G channel, no useful telluric absorption features exist. Instead, we measured the centroid offset of the dominant [O I] 5577\,\AA\,emission line and applied the resulting flexure correction uniformly across the channel (see Section~\ref{sec:sky_emission}). 

We plot example flexure correction curves derived from telluric absorption features with black lines in the left column of Figure~\ref{fig:flexure_curves}. The blue points show measured telluric shifts, while the orange crosses show where the flexure curve is evaluated to derive corrections for local RV measurements (see Section~\ref{sec:radial_velocities}). We find that the best-fit flexure correction can be strongly wavelength-dependent, especially in the I channel, where it varies by tens of km s$^{-1}$ over the spectral range. 

\subsubsection{Sky Emission Lines}
\label{sec:sky_emission}

In case telluric absorption features are weak or unavailable, sky emission lines with known wavelengths can be used to measure flexure corrections instead. To find anchors for the flexure curve, we selected strong emission lines in the G, R, or I channels (i.e., with central peak flux $> 2.0, 5.0,$ or $30.0 \times 10^{-16}$ erg s$^{-1}$ cm$^{-2}$ \AA$^{-1}$ respectively) from the atlas of \citet{hanuschik_emission_2003}. We modeled the emission lines as Gaussian profiles superimposed on a linearly varying local continuum:

\begin{equation}
F_{\rm sky}(\lambda) \simeq b_0+b_1(\lambda-\lambda_0)+A\exp\left[-\frac{(\lambda-\lambda_c)^2}{2\sigma_\lambda^2}\right].
\end{equation}

Here, $F_{\rm sky}$ is the 1D sky model extracted at the location of the science trace, $b_0$ and $b_1$ describe the local linear continuum, $A$ is the line amplitude, $\lambda_c$ is the fitted centroid, and $\sigma_{\lambda}$ is the line width. Given the tabulated emission-line wavelength $\lambda_{\rm ref}$, the line centroid offset was then converted to a flexure correction in velocity space:

$$
v_{\rm flex}(\lambda_c) = c\,\frac{\lambda_c - \lambda_{\rm ref}}{\lambda_{\rm ref}}.
$$

\begin{figure}[h]
    \centering
    \includegraphics[width=\columnwidth]{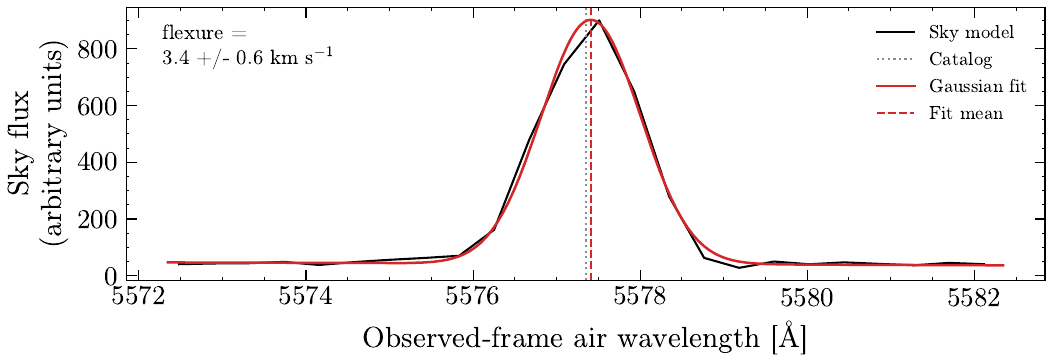}
    \caption{Example of flexure measurement using the [O I] 5577\,\AA~sky emission line in the G channel. We show a Gaussian fit to the observed line profile (black) in red. Based on the centroid shift relative to the \citet{hanuschik_emission_2003} catalog, the measured flexure is $3.4 \pm 0.6$ km s$^{-1}$.}
    \label{fig:sky_flexure}
\end{figure}

As an example, we show a Gaussian fit to the forbidden [O I] sky emission line at air wavelength 5577.34\,\AA~in Figure~\ref{fig:sky_flexure}. Based on the mean of the fitted Gaussian, the measured centroid shift translates to a flexure of $3.4 \pm 0.6$ km s$^{-1}$. We used this measurement to correct the stellar RV measured in the G channel.

To avoid close blends or crowded forests in the R or I channels, we enforced a minimum separation of 10\,\AA~between sky emission lines selected as anchors. We fitted the Gaussian model in a $\pm 5$\,\AA~window around the reference wavelength given in \citet{hanuschik_emission_2003}. We accepted an emission line anchor as valid if the fit converged and the flexure uncertainty was $< 35$ km s$^{-1}$. We performed $4\sigma$ clipping to remove extreme outliers while preserving smooth wavelength-dependent structure, and linearly interpolated between accepted anchors to derive the final flexure curve. 

\begin{figure*}
    \centering
    \includegraphics[width=\textwidth]{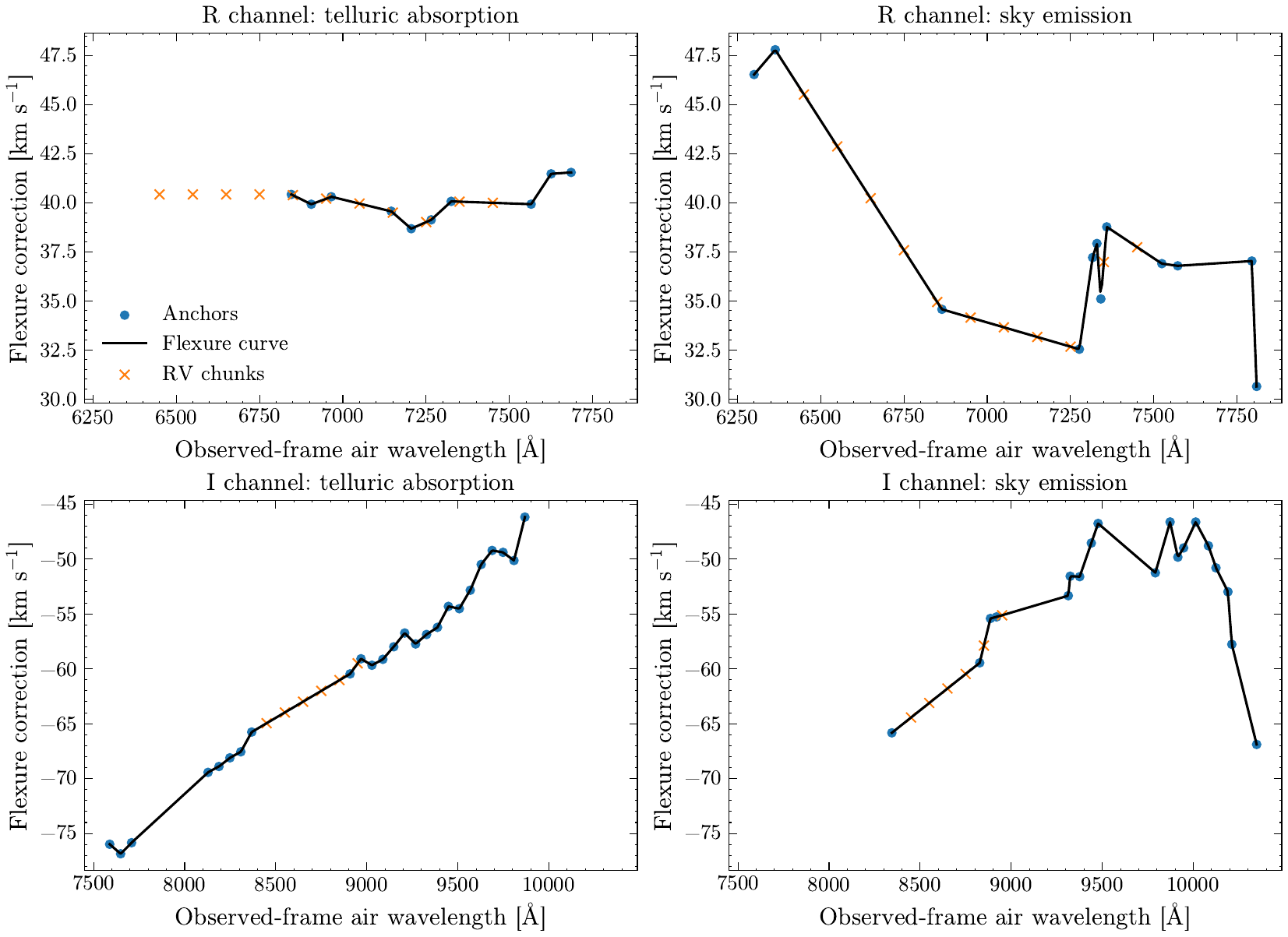}
    \caption{Example flexure correction curves derived using telluric absorption (left column) and emission (right column) features in the R and I channels. Blue points represent the local flexure measurements, black lines show the interpolated flexure curves, and orange crosses show where the flexure curve is evaluated to derive local flexure corrections for stellar RV measurements. In the R channel, the curve derived from telluric absorption features is more stable than the one derived from strong, isolated sky emission lines. In the I channel, where telluric features are strongest, the curves evaluate to similar values in the overlapping wavelength range.}
    \label{fig:flexure_curves}
\end{figure*}

We show examples of flexure correction curves derived from sky emission lines in the right column of Figure~\ref{fig:flexure_curves}. Comparing the flexure correction curves derived in the R channel, we find that the sky emission anchors display larger variability than the telluric absorption anchors. On the other hand, in the I channel (where telluric features are strongest), the curves evaluate to similar values in the overlapping wavelength range. In Section~\ref{sec:results}, we find that telluric absorption features provide a more reliable flexure correction than sky emission lines.

\subsection{Radial Velocities}
\label{sec:radial_velocities}

We measured stellar RVs in each channel via comparison with synthetic template spectra. Specifically, for each target, we chose the template spectrum as the BOSZ \citep{Kurucz1979, Kurucz1993, bosz_2024} spectrum with spectroscopic parameters closest to those reported in the Gaia-ESO database \citep{randich_eso_2022}. The template spectra used cover a grid of metallicity $\rm [M/H] \in [-0.5, +0.5]$ spaced by $0.25$, effective temperature $T_{\text{eff}} \in [4000 \text{ K}, 6000 \text{ K}]$ spaced by $250$ K, and surface gravity $\log{\left(g / \text{cm s$^{-2}$}\right)} \in [1.5 , 4.5]$ spaced by $0.5$. We assumed no rotational broadening. Instead, we used a Gaussian kernel to broaden the template (originally at $R = 50,000$) to the median instrumental resolving power in the given channel, estimated from the full-width half-maxima of strong, isolated sky emission lines (see Section~\ref{sec:campaign}).

To robustly determine the stellar RV, we broke the observed spectrum into wavelength ``chunks'' of equal length. By default, we used 100\,\AA~chunks from 6400--7500\,\AA~in the R channel, and from 8400--9000\,\AA~in the I channel. Since only one flexure measurement was made in the G channel (see Section~\ref{sec:sky_emission}), we did not break it into chunks, and instead used the entire $5150$--$5700$\,\AA~wavelength range. We normalized each chunk by dividing by a local pseudo-continuum estimated with a 21\,\AA~median filter. Pixels where the shifted telluric model predicts strong atmospheric absorption (i.e., transmission $<97\%$) were masked. Then, within each chunk, we determined the RV shift of the template that minimized the $\chi^2$ statistic with respect to the observed spectrum:

\begin{equation}
\label{eq:chi_squared}
\chi^2(v)=\sum_i\frac{\left[f_i-S_i(v)\right]^2}{\sigma_i^2},
\end{equation}

\noindent where $f_i$ is the normalized observed flux in pixel $i$, $S_i(v)$ is the flux of the stellar template shifted by trial velocity $v$ at that pixel, and $\sigma_i$ is the flux uncertainty. We initially estimated $\sigma_i$ as the inverse of the measured SNR (see Section~\ref{sec:extraction}). After finding the best-fit RV shift, we replaced this estimate with the sigma-clipped standard deviation of the flux residuals at continuum pixels (i.e., as predicted by the template), in order to account for variations in the effective noise level between chunks. The error on the best-fit RV shift was based on the width of the minimum, defined as the velocity interval over which $\chi^2(v)$ changes by one.

Let $\lambda_{\rm chunk}$ be the wavelength at the center of a given spectral chunk. To correct for flexure, the derived flexure curve from Section~\ref{sec:flexure_corrections} was evaluated at this wavelength, producing a shift $v_{\rm flex}({\lambda_{\rm chunk}})$, which we subtracted from the RV shift obtained from Equation~\ref{eq:chi_squared}. The best-fit RV in an individual chunk was also corrected for the Earth's barycenter motion $v_{\rm bary}$ computed at the mid-exposure time of the observation, as the NGPS QuickLook DRP does not apply this correction by default. We used the Astropy package \citep{2013A&A...558A..33A, 2018AJ....156..123A, 2022ApJ...935..167A} to compute the barycentric correction. In that convention, this correction is added to the observed RV:

$$
v_{\rm chunk} = \underset{v}{\arg\min}\;\chi^2(v) - v_{\rm flex}(\lambda_{\rm chunk}) + v_{\rm bary}.
$$

Finally, the individual chunk RVs were robustly combined after quality cuts and outlier rejection to produce a final stellar RV measurement $v_{\star}$. In detail, we rejected chunks that were dominated by telluric features, had $\chi^2$ curves without a well-defined minimum, or had large RV errors $\sigma_{v_{\rm chunk}}> 25$ km s$^{-1}$. In the G channel, which was not broken into chunks (and thus lacked the redundancy of the R and I channels), we additionally rejected spectra that showed an unusually large fraction of extreme data-model residuals (i.e., absolute normalized residual $> 2$), which could have arisen due to poor sky subtraction, cosmic rays, or detector artifacts. In the R and I channels, we used $3\sigma$ clipping to identify and remove chunk RVs that are outliers, and combined the remaining measurements using a weighted average, with weights $w_{\rm chunk}$ given by:

\begin{equation}
    w_{\rm chunk} = \frac{1}{\sigma_{v_{\rm chunk}}^2 + \sigma_{\rm floor}^2}.
\end{equation} 

Here, a $\sigma_{\rm floor} = 3$ km s$^{-1}$ noise floor was used to avoid any single chunk dominating the final RV measurement. To account for both the RV uncertainty in each chunk and the scatter in RV measurements between chunks, we computed the non-flexure uncertainty in the final RV measurement as:

\begin{equation}
     \sigma^2_{v_{\rm chunks}} = \max\left[\left(\sum_{\rm chunks} w_{\rm chunk}\right)^{-1},\, \frac{\hat{\sigma}^2_{\rm chunks}}{N_{\rm chunks}} \right],
\end{equation}

\noindent where $N_{\rm chunks}$ is the number of combined chunks and $\hat{\sigma}_{\rm chunks}$ is the standard deviation of the chunk RVs estimated from their median absolute deviation \citep[e.g.,][]{stoehr_dersnr_2008}:

\begin{equation}
   \hat{\sigma}_{\rm chunks} = 1.4826 \operatorname{Mdn}\left(
\left|v_{\rm chunk}-\operatorname{Mdn}\left(v_{\rm chunks}\right)\right|
\right).
\end{equation}

We estimated the uncertainty introduced by the flexure correction as the standard deviation $\sigma_{v_{\rm flex}}$ of the final RV measurement over 1000 Monte Carlo realizations of the flexure correction curve. Specifically, for each draw, we perturbed the local flexure measurements, reconstructed the interpolated flexure curve, re-evaluated it at each chunk, and re-combined the chunk RVs. The total RV uncertainty $\sigma_{v_*}$ was computed by adding  $\sigma_{v_{\rm chunks}}$ and $\sigma_{v_{\rm flex}}$ in quadrature:

\begin{equation}
  \sigma_{v_*} = \sqrt{\sigma^2_{v_{\rm chunks}} + \sigma_{v_{\rm flex}}^2}.
\end{equation}

Our prescription imposes a lower bound of $3/\sqrt{N_{\rm chunks}}$ km s$^{-1}$ on the total uncertainty, but the reported RV uncertainty need not approach this value at arbitrarily large SNR, because model-observation mismatch and calibration errors can sustain chunk-to-chunk scatter and uncertainty in the fitted flexure corrections.

\begin{figure*}
    \centering
    \includegraphics[width=\textwidth]{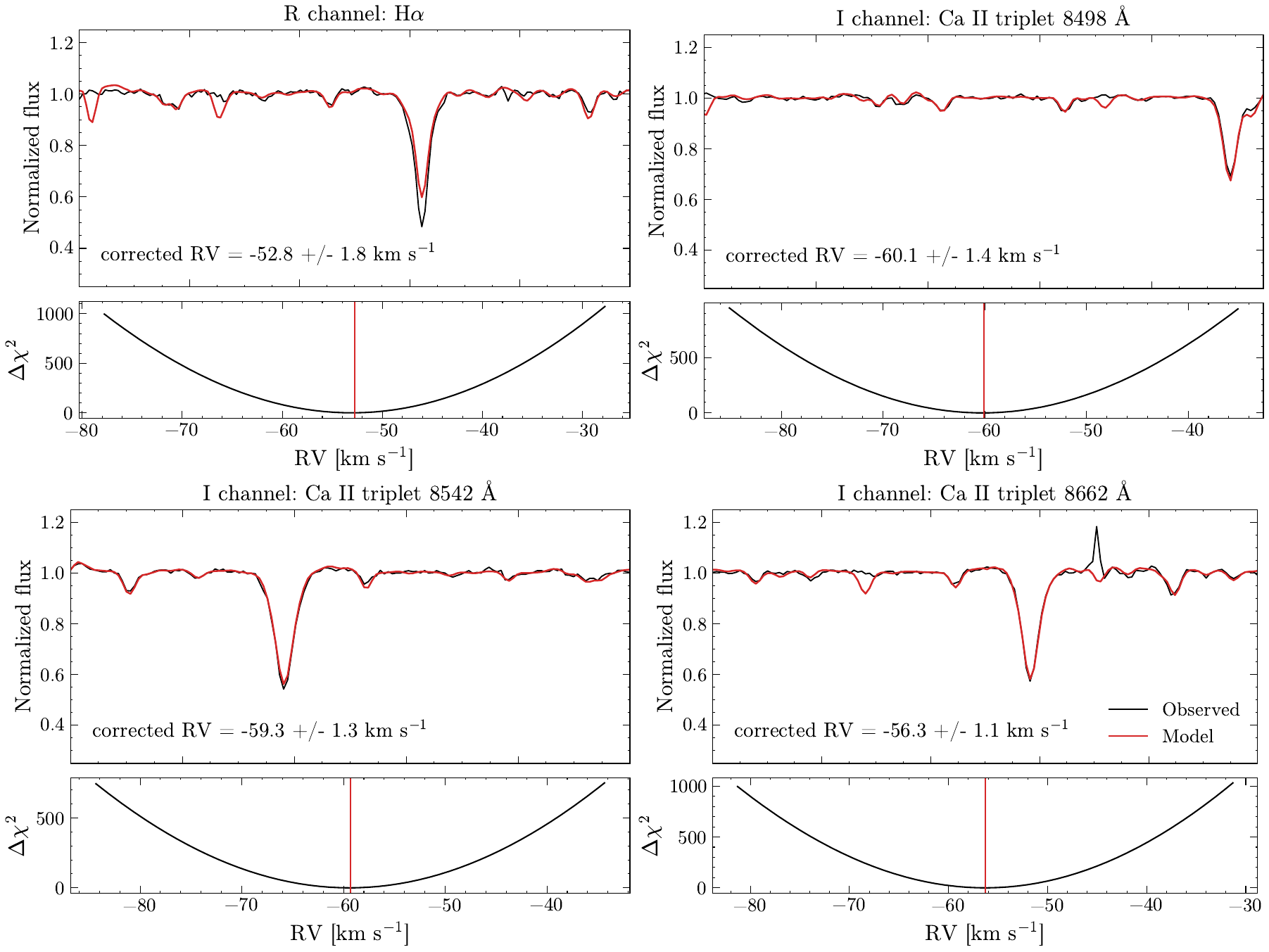}
    \caption{Comparison of observed spectrum and shifted BOSZ stellar template for an example RV standard star, focusing on the RV signal from the H$\alpha$ absorption line in the R channel and the Ca II triplet in the I channel. After combining the RV signal from several spectral ``chunks'' (10 in the R channel and 5 in the I channel) and correcting for flexure and barycentric motion, the final measured RV is $-58.0 \pm 4.2$ km s$^{-1}$ in the R channel and $-59.0 \pm 1.6$ km s$^{-1}$ in the I channel.}
    \label{fig:rv_measurement}
\end{figure*}

We compare the observed spectrum and shifted BOSZ stellar template for an example RV standard star in Figure~\ref{fig:rv_measurement}, focusing on wavelength chunks containing H$\alpha$ (in the R channel) and the Ca II triplet (in the I channel). In these chunks, there are no telluric-rejected pixels, implying that the observed spectra are dominated by stellar features. The $\chi^2$ curves corresponding to the overlays are shown as well, with the annotations reporting the best-fit chunk RVs after applying the flexure and barycentric corrections. After robustly combining all chunk RVs, the final measured RV is $-58.0 \pm 4.2$ km~s$^{-1}$ in the R channel and $-59.0 \pm 1.6$ km~s$^{-1}$ in the I channel. These are consistent with the reported Gaia-ESO RV of $-60.32$ km s$^{-1}$ to within $1\sigma$.

\section{Results}
\label{sec:results}

\subsection{Fiducial Pipeline}

\begin{figure*}
    \centering
    \includegraphics[width=\textwidth]{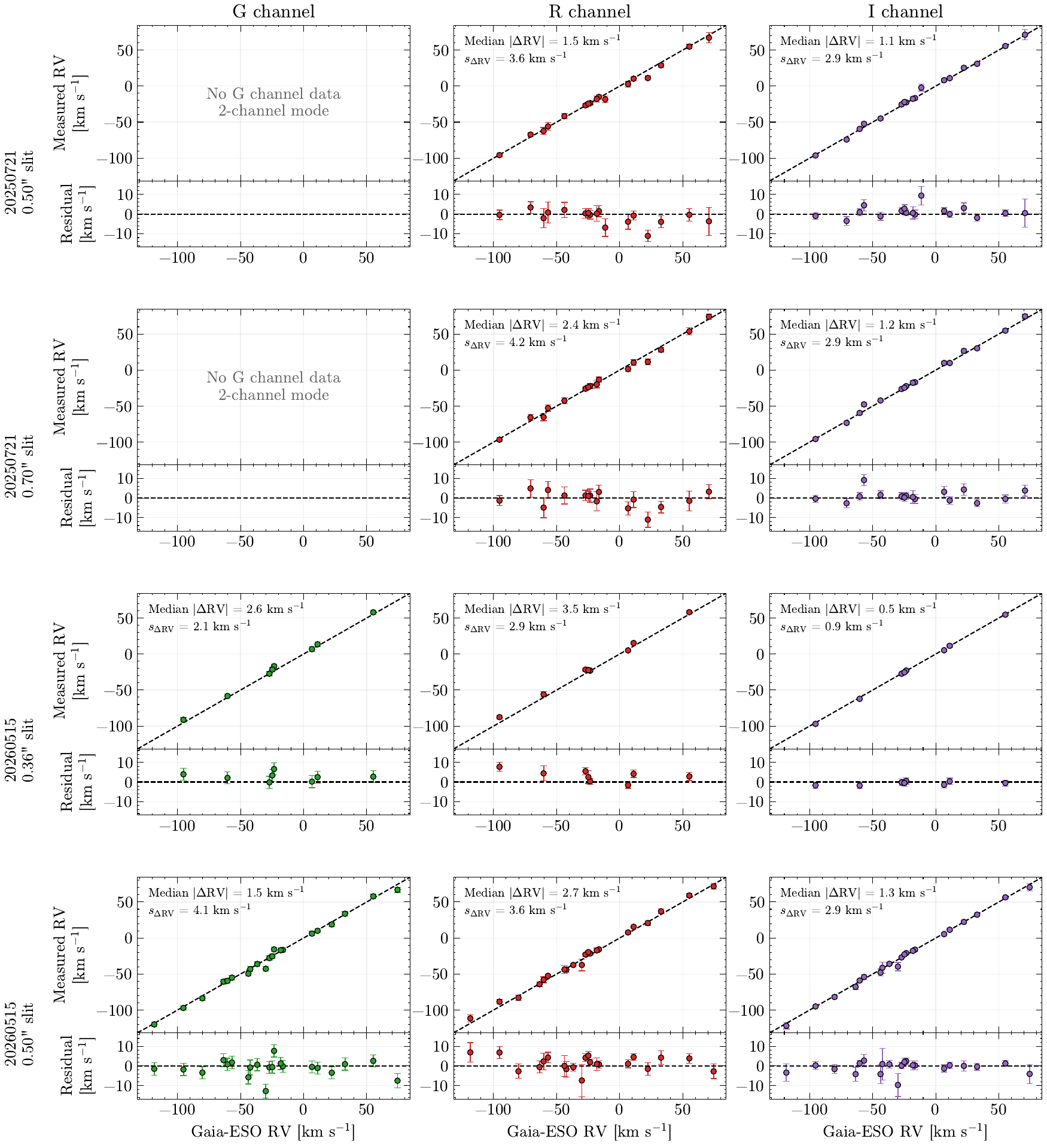}
    \caption{Comparison of flexure-corrected RVs measured with our pipeline to catalog RVs for all Gaia-ESO RV standards observed during our NGPS campaign. Each column shows measurements in a given NGPS channel, while each row corresponds to a specific combination of observing night and slit width. We find that the median absolute RV residual ranges from $\approx 0.5$ to $\approx 3.5$ km s$^{-1}$, with 88\% of all 153 measurements within 5 km s$^{-1}$ of the catalog values. The I channel generally gives the best accuracy and precision, but we find no clear improvement with decreasing slit width.}
    \label{fig:full_comparison}
\end{figure*}

We ran our RV measurement pipeline on all 153 \mbox{G-,} R-, and I-channel spectra of Gaia-ESO RV standards obtained during our NGPS campaign.\footnote{We do not analyze the U band spectra, because that channel contains neither strong telluric absorption features nor useful sky-emission lines for measuring flexure. Furthermore, all 153 analyzed G-, R-, and I-channel spectra have SNR $> 5$, while several U-channel spectra do not.} We used the same defaults as described in Section~\ref{sec:flexure_corrections}: flexure was measured from the [O I] 5577\,\AA~sky emission line in the G channel, and from telluric absorption features in the R and I channels. We show all of our flexure-corrected stellar RVs in Figure~\ref{fig:full_comparison}. Each column corresponds to an NGPS channel and each row corresponds to a combination of a slit width and observing night. In every subplot, the upper panel compares the measured RV to the Gaia-ESO catalog RV, while the lower panel shows the residual relative to the catalog value. The upper panel also reports the median absolute RV residual and sample standard deviation for that batch test. In the residual panels, the error bars represent the uncertainties in the pipeline RV and the Gaia-ESO catalog RV added in quadrature. We find that, across combinations of channel and slit width, the median absolute residual of the measured RVs relative to the Gaia-ESO RVs ranges from $\approx 0.5$ km s$^{-1}$ to $\approx 3.5$ km s$^{-1}$, with RV residuals being modestly larger for fainter targets than brighter targets. 

\subsection{No Flexure Corrections}
\label{sec:no_corr}

\begin{figure*}
    \centering
    \includegraphics[width=\textwidth]{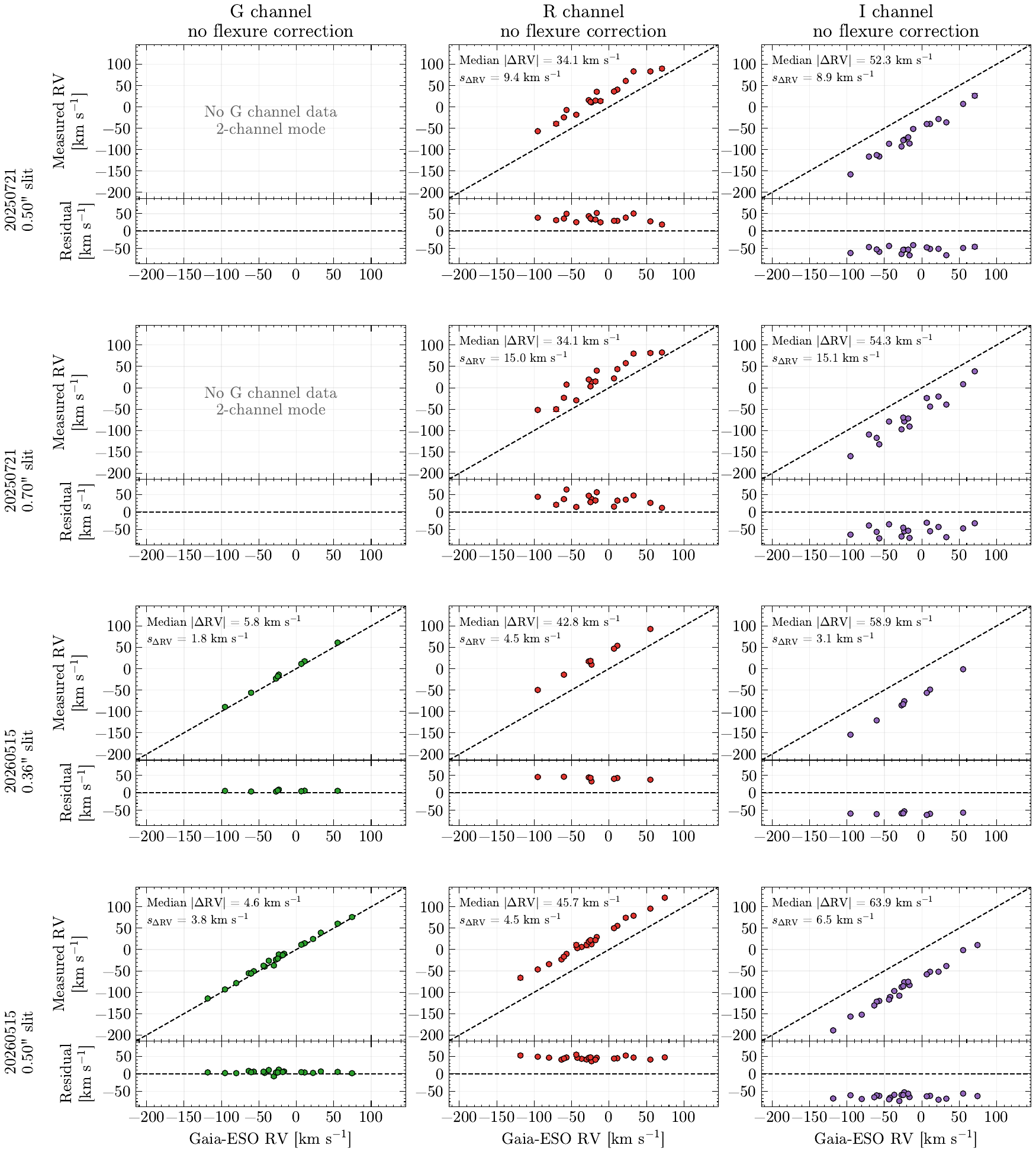}
    \caption{Same as Figure~\ref{fig:full_comparison}, but without any flexure corrections. The derived RVs show large systematic shifts relative to the Gaia-ESO catalog values, especially in the R and I channels, demonstrating the necessity of flexure measurements for obtaining accurate RVs.}
    \label{fig:no_corr}
\end{figure*}

To demonstrate the importance of flexure corrections, we repeated the batch validation test without them. We show the results in Figure~\ref{fig:no_corr}. In the R and I channels, the derived RVs show large systematic offsets: depending on night and slit width, the mean RV residual ranges from $+34.4$ to $+45.2$ km s$^{-1}$ in R and from $-65.2$ to $-53.0$ km s$^{-1}$ in I, with no measurement recovered to within $10$ km s$^{-1}$ of the Gaia-ESO value. The degradation in performance is more modest in G, where the mean offset is approximately $+4.8$ km s$^{-1}$. Across all channels, omitting the flexure correction increases the median absolute RV residual from $1.59$ to $44.30$ km s$^{-1}$, decreases the fraction recovered within 5 km s$^{-1}$ from $88$\% to $10$\%, and increases the reduced chi-squared from $\approx 1.3$ to $\approx 608$. This exercise demonstrates that (wavelength-dependent) flexure corrections are essential for obtaining accurate RVs with NGPS, particularly in the R and I channels. In addition, although the RV offsets are correlated across targets, they shift both over the course of a night and between nights. Thus, it is insufficient to determine a flexure correction based on a single RV standard and apply that correction globally to all RV measurements.

\subsection{Sky Emission Fallback}

\begin{figure*}
    \centering
    \includegraphics[width=\textwidth]{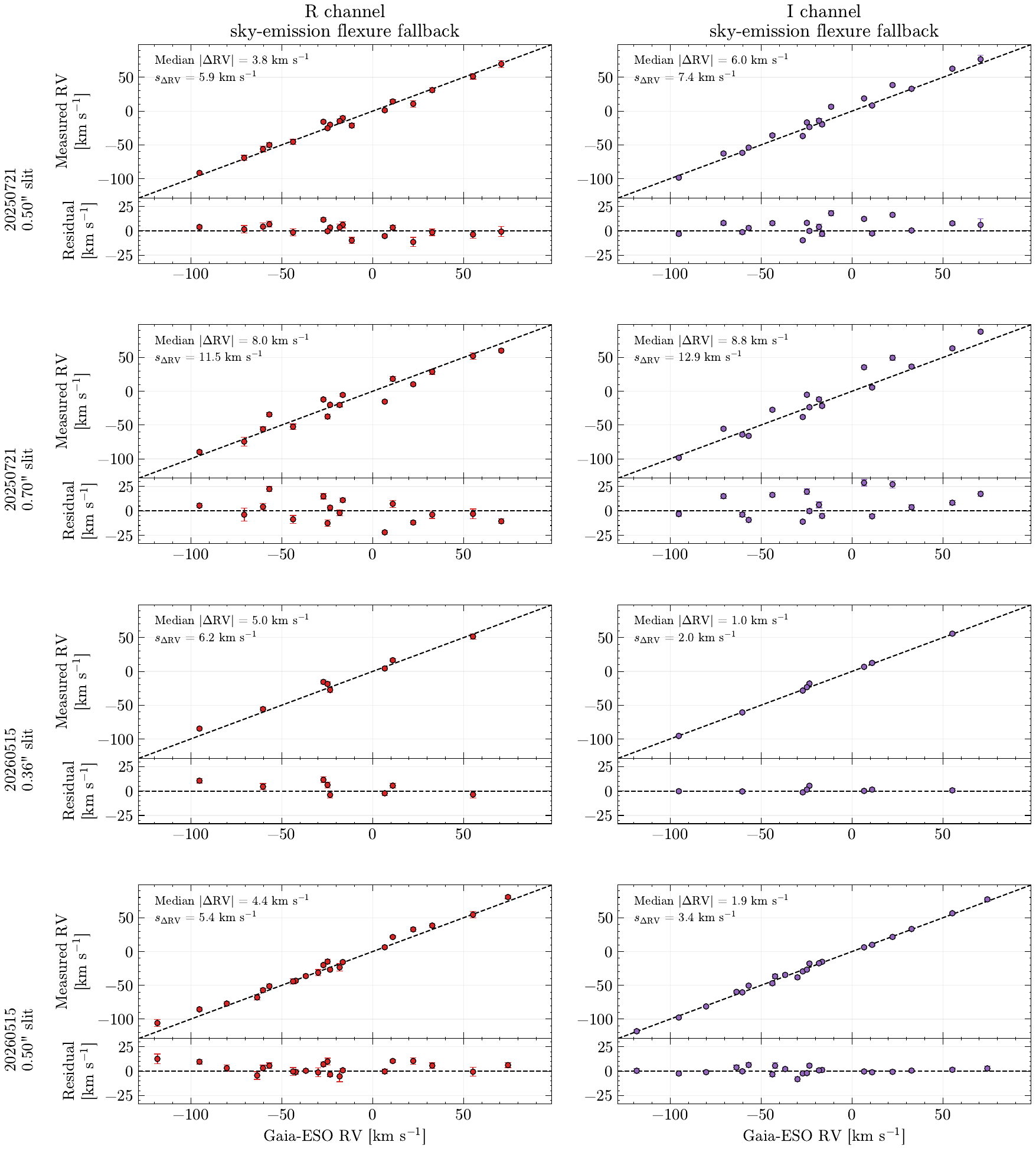}
    \caption{Same as Figure~\ref{fig:full_comparison}, but using sky emission lines as a fallback option to measure flexure. We find that the median absolute RV residual (relative to the Gaia-ESO catalog value) increases noticeably, demonstrating that telluric absorption features provide a more reliable method of flexure correction in the R and I channels.}
    \label{fig:batch_emission}
\end{figure*}

As a fallback option, it is also possible to use strong sky emission lines as a flexure source in the R and I channels. We re-ran our pipeline with this option and show the derived best-fit RVs in Figure~\ref{fig:batch_emission}. We find that the RV performance worsens considerably in this case, with the median absolute RV residual reaching up to $\approx8.8$~km~s$^{-1}$ for the 0.70$\arcsec$ slit in the I channel.

\subsection{Summary}

\begin{deluxetable*}{ccccccccc}
\tablecaption{
Accuracy and precision metrics for samples of RV standard stars categorized by night, slit width, and channel. In each row, $N$ is the number of successful RV
measurements, $\Delta{\rm RV} \equiv {\rm RV}_{\rm NGPS}-{\rm RV}_{\rm Gaia\mbox{-}ESO}$, and $s_{\Delta{\rm RV}}$ is the corresponding sample standard deviation. The uncertainty on $\langle\Delta{\rm RV}\rangle$ is the standard error of the mean. The reported $\chi^2_{\nu}$ includes the pipeline's RV uncertainty and
Gaia-ESO RV uncertainty added in quadrature. Finally, $f_{5\,{\rm km\,s^{-1}}}$ and $f_{10\,{\rm km\,s^{-1}}}$ refer to the fraction of measurements satisfying $|\Delta{\rm RV}|\leq5~{\rm km\,s^{-1}}$ and $\leq10~{\rm km\,s^{-1}}$, respectively.
\label{tab:metrics}
}
\tablehead{
\colhead{UT date/slit} &
\colhead{Channel} &
\colhead{$N$} &
\colhead{$\langle\Delta{\rm RV}\rangle$} &
\colhead{$s_{\Delta{\rm RV}}$} &
\colhead{Median $\lvert \Delta{\rm RV} \rvert$} &
\colhead{$\chi_\nu^2$} &
\colhead{$f_{5\,{\rm km\,s^{-1}}}$} &
\colhead{$f_{10\,{\rm km\,s^{-1}}}$} \\
\colhead{} &
\colhead{} &
\colhead{} &
\colhead{(km s$^{-1}$)} &
\colhead{(km s$^{-1}$)} &
\colhead{(km s$^{-1}$)} &
\colhead{} &
\colhead{} &
\colhead{}
}
\startdata
\sidehead{\textbf{Fiducial pipeline}}
2025 Jul 21, $0.50\arcsec$ & R & 17 &
$-1.46\pm0.87$ & 3.57 & 1.48 & 1.37 & 88\% & 94\% \\
 & I & 17 &
$+1.05\pm0.69$ & 2.86 & 1.13 & 0.97 & 94\% & 100\% \\
2025 Jul 21, $0.70\arcsec$ & R & 16 &
$-0.68\pm1.04$ & 4.16 & 2.40 & 1.15 & 88\% & 94\% \\
 & I & 16 &
$+1.11\pm0.73$ & 2.92 & 1.19 & 1.24 & 94\% & 100\% \\
2026 May 15, $0.36\arcsec$ & G & 8 &
$+2.64\pm0.76$ & 2.15 & 2.60 & 1.17 & 88\% & 100\% \\
 & R & 8 &
$+3.24\pm1.03$ & 2.91 & 3.50 & 3.65 & 75\% & 100\% \\
 & I & 8 &
$-0.67\pm0.32$ & 0.90 & 0.54 & 0.41 & 100\% & 100\% \\
2026 May 15, $0.50\arcsec$ & G & 21 &
$-1.02\pm0.91$ & 4.15 & 1.46 & 1.56 & 81\% & 95\% \\
 & R & 21 &
$+1.44\pm0.78$ & 3.57 & 2.70 & 1.73 & 81\% & 100\% \\
 & I & 21 &
$-0.73\pm0.64$ & 2.93 & 1.27 & 0.62 & 95\% & 100\% \\
\hline
Overall & G, R, I & 153 &
$+0.23\pm0.28$ & 3.52 & 1.59 & 1.32 & 88\% & 98\% \\
\hline
\sidehead{\textbf{Sky-emission fallback}}
2025 Jul 21, $0.50\arcsec$ & R & 17 &
$+0.48\pm1.43$ & 5.89 & 3.80 & 4.41 & 65\% & 88\% \\
 & I & 17 &
$+4.14\pm1.80$ & 7.41 & 6.01 & 16.14 & 47\% & 82\% \\
2025 Jul 21, $0.70\arcsec$ & R & 16 &
$-0.63\pm2.86$ & 11.46 & 7.97 & 21.77 & 38\% & 56\% \\
 & I & 16 &
$+6.56\pm3.22$ & 12.88 & 8.80 & 25.90 & 25\% & 56\% \\
2026 May 15, $0.36\arcsec$ & R & 8 &
$+3.59\pm2.18$ & 6.16 & 5.01 & 6.52 & 50\% & 75\% \\
 & I & 8 &
$+0.92\pm0.72$ & 2.05 & 1.01 & 1.85 & 88\% & 100\% \\
2026 May 15, $0.50\arcsec$ & R & 21 &
$+3.29\pm1.17$ & 5.38 & 4.40 & 5.42 & 52\% & 86\% \\
 & I & 21 &
$+0.48\pm0.74$ & 3.37 & 1.94 & 2.36 & 81\% & 100\% \\
\hline
Overall & R, I & 124 &
$+2.33\pm0.71$ & 7.89 & 3.88 & 10.82 & 55\% & 81\% \\
\enddata
\end{deluxetable*}

We summarize accuracy and precision metrics for the validation tests in Figures~\ref{fig:full_comparison} and \ref{fig:batch_emission} in Table~\ref{tab:metrics}. For each combination of night, slit width, and channel, we report the number of successful RV measurements (i.e., that pass the aforementioned quality criteria), average RV residual relative to the Gaia-ESO catalog value (and its standard error), sample standard deviation, median absolute residual, and reduced chi-squared value. We also report the fraction of measurements with RV residual smaller than $5$ and $10$ km s$^{-1}$. We find that, for the fiducial pipeline, the overall mean RV residual is $+0.23\pm0.28$~km~s$^{-1}$, the standard deviation of the residuals is $3.52$~km~s$^{-1}$, and 88\% and 98\% of the 153 measurements lie within 5 and 10~km~s$^{-1}$ of the catalog RVs, respectively. The overall reduced chi-squared of $\chi_\nu^2=1.32$ indicates that the propagated RV uncertainties reproduce the observed scatter reasonably well, though they may remain slightly underestimated.

When sky emission is used as a fallback option for flexure measurement in the R and I channels, accuracy and precision worsen significantly. Overall, the average RV residual increases to $+2.33 \pm 0.71$ km~s$^{-1}$, and the standard deviation of the residuals increases to $7.89$ km~s$^{-1}$. Only 55\% and 81\% of the 124 measurements lie within 5 and 10~km~s$^{-1}$ of the catalog RVs, respectively, while the overall reduced chi-squared of $\chi_\nu^2=10.82$ is driven by large outliers. This comparison reinforces the fact that telluric absorption should be preferred for flexure measurements whenever possible --- such features are imprinted directly on the spectrum of the illuminating point source, whereas sky emission fills the entire slit and does not track miscentering in the same way.

Comparing paired results for targets observed with the $0.5''$ slit in the R and I channels in both 2025 and 2026, we find that the measured RVs in the R channel shift by $+4.0 \pm 0.9$ km s$^{-1}$ on average between epochs, while the measured RVs in the I channel shift negligibly. This may be due to the fact that the flexure correction evaluated at H$\alpha$ is clamped to the value of the bluest accepted anchor, and therefore less stable than flexure corrections evaluated around the Ca II triplet, which are well-constrained by surrounding anchors. Indeed, the measured RVs tend to be the most accurate and precise in the I channel. Nevertheless, the derived RVs tend to agree well across channels; if we combine the available channel measurements for each observation with an inverse-variance-weighted average, we find that $\approx 97\%$ of the resulting RVs are consistent with the Gaia-ESO catalog values to within $5$ km s$^{-1}$.

\begin{figure*}
    \centering
    \includegraphics[width=\textwidth]{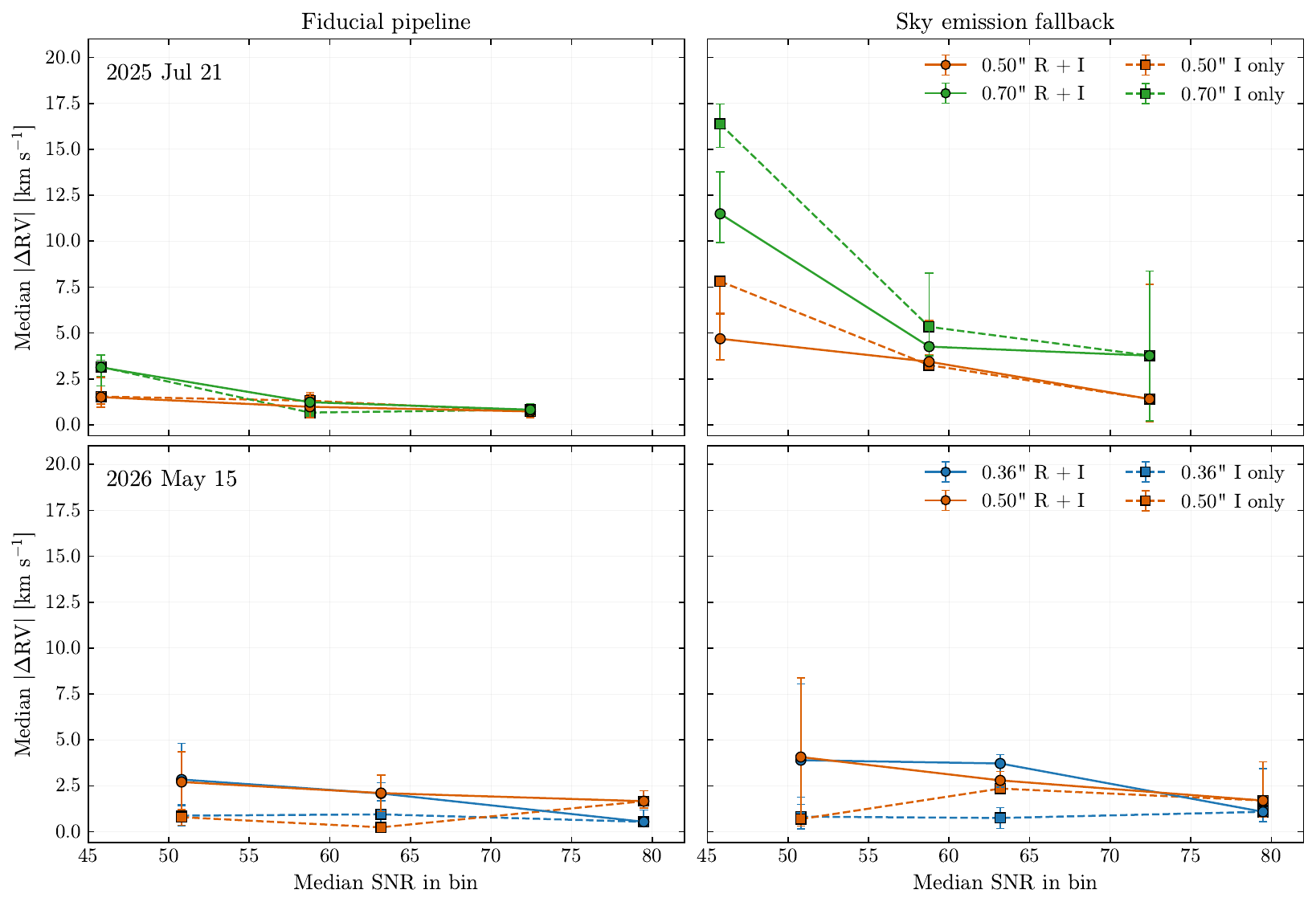}
    \caption{Effect of slit width on the median absolute RV residual as a function of SNR. From a paired comparison of spectra taken in the R and I channels (solid lines), we find that the improvement in median absolute RV residual is statistically significant from the $0.70''$ slit to the $0.50''$ slit, but not from the $0.50''$ slit to the $0.36''$ slit. This holds true regardless of whether flexure is measured from telluric absorption features or sky emission lines. On the other hand, when restricting the paired comparison to spectra taken only in the I channel (dashed lines), the fiducial pipeline's improvement from the $0.70''$ slit to the $0.50''$ slit is no longer statistically significant.}
    \label{fig:snr_slit_width}
\end{figure*}

We investigate the variation of median absolute RV residual with SNR and slit width in Figure~\ref{fig:snr_slit_width}. Specifically, we group spectra from the R and I channels into SNR bins and compare the performance of different slit widths for the same targets on each night. The error bars are 16th-84th percentile bootstrap intervals. We find a statistically significant decrease in the median $\lvert \Delta \text{RV}\rvert$ from the $0.70''$ slit to the $0.50''$ slit in 2025, with an improvement of $0.66^{+0.50}_{-0.22}$ km s$^{-1}$ for the fiducial pipeline and $3.10^{+4.07}_{-1.49}$ km s$^{-1}$ for the sky emission fallback option. Here, the uncertainties represent $95\%$ confidence intervals. On the other hand, we do not find a statistically significant decrease in the median $\lvert \Delta \text{RV}\rvert$ from the $0.50''$ slit to the $0.36''$ slit in 2026. Furthermore, when restricting the analysis to the I channel (where RV performance tends to be the best), we find that the fiducial pipeline's improvement from the $0.7\arcsec$ slit to the $0.5\arcsec$ slit in 2025 is no longer statistically significant. We conclude that spectral resolving power increases modestly with decreasing slit width, but there is mixed evidence for a corresponding improvement in RV performance. 

\section{Discussion} 
\label{sec:discussion}

\subsection{Using the Pipeline}

The source code for our RV measurement pipeline, and a tutorial notebook demonstrating its use, is publicly available.\footnote{\url{https://github.com/pranav-nagarajan/NGPS-RV-Calibration}} We also provide a command-line interface (CLI) to run the pipeline. The CLI accepts as user input a QuickLook DRP reduced NGPS 2D spectrum in a given channel, a stellar template spectrum, and several optional arguments. The user can specify an RV prior and a restricted wavelength range, provide custom flux uncertainties or sky emission line catalogs, and choose the source for flexure measurements in the R or I channels. The pipeline returns the best-fit RV and corresponding uncertainty, along with summary plots and tables describing the derived flexure correction curve and fitted spectral chunks.

It is important to provide an appropriate template spectrum for the science target of interest. In this work, we do not investigate stars that are cooler than 4000 K, hotter than 6000 K, metal-poor, or fast-rotating (in doing so, we minimize template mismatch with the adopted grid of BOSZ model spectra). We also do not investigate stars that have evolved beyond the red giant branch. Finally, we do not investigate double-lined spectroscopic binaries, which would require a composite template spectrum. 

Our results characterize the uncertainty associated with the flexure correction. The total RV uncertainty may be larger for stars of spectral types with fewer or weaker absorption features. However, at comparable SNR and within a given channel, the reported uncertainty in the flexure correction should remain similar to that measured for our sample.

\subsection{Practical Recommendations}

Based on the results of our batch validation tests in Section~\ref{sec:results}, we find that the I channel provides the most consistently reliable RV measurements, while the R and G channels provide useful independent consistency checks. Notably, the I channel features the Ca II triplet for relatively cool stars and Paschen lines for relatively hot stars, making it useful over a broad range of spectral types. Furthermore, we find that combining all available channel measurements with an inverse-variance-weighted average increases the fraction of observations whose RV is recovered to within $5$ km s$^{-1}$ of the corresponding Gaia-ESO catalog value. This advantage persists for the subset of observations in which at least one pair of channel RV measurements differs by more than its combined $1\sigma$ uncertainty. Such disagreement should nevertheless prompt inspection of the flexure curves, fitted spectral chunks, and other pipeline diagnostics.

For RV measurements using telluric absorption as a flexure source, we recommend retaining the $1\sigma$ RV uncertainty returned by the pipeline within the SNR regime ($\approx 15$--$100$) tested here. The median uncertainty for our R- and I-channel sample is $\approx 2.4$ km s$^{-1}$, and about 68\% of these RV measurements agree with the Gaia-ESO RVs within their combined $1\sigma$ uncertainties. When sky emission is used instead, the returned RV uncertainty tends to substantially underestimate the observed scatter. We therefore recommend treating such measurements as preliminary, and adding an empirical systematic uncertainty of $\approx 7$ km s$^{-1}$ in quadrature. Whenever possible, we recommend using telluric absorption features to measure flexure corrections. Since sky emission fills the slit instead of tracking the point source, sky emission lines should only be used as an alternate flexure source when deriving preliminary RV measurements. It is imperative to correct for flexure to measure accurate RVs with NGPS, with large systematic offsets arising when such corrections are omitted (see Figure~\ref{fig:no_corr} and Section~\ref{sec:no_corr}). Leveraging the telluric information inherent in the observations themselves saves time relative to arc calibrations taken throughout the night.

We do not find a strong trend in median absolute RV residual with slit width, though our sample size of targets observed with the narrowest slit width of $0.36''$ is small. While any slit width $\gtrsim 0.36\arcsec$ and $\lesssim 0.7\arcsec$ will preserve the RV performance of NGPS, there is a trade-off between transmitted light and spectral resolution. We recommend using a $0.5''$ slit by default, with a wider slit appropriate for faint sources, and a narrower slit appropriate for resolving blended features in bright sources. While observers may wish to use $1.5''$ or $2.0''$ slits for e.g., transient or extended sources, we only tested slit widths of up to $0.70''$ in this work.

We find that the NGPS wavelength solution varies from night to night. While our pipeline accounts for this, measured RVs in the R channel still shift on average by $+4.0 \pm 0.9$ km s$^{-1}$ between epochs. On the other hand, RVs measured in the I channel are stable. This is likely because the flexure correction at H$\alpha$ is clamped to the value of the bluest anchor in the R channel, while the flexure correction at the Ca II triplet is interpolated between anchors in the I channel. In the G channel, a correction based on a single prominent emission line seems to work well, suggesting that flexure in that channel is currently not very wavelength-dependent (see also Section~\ref{sec:no_corr}). On the other hand, we were unable to test whether this remains true on months-long timescales, since NGPS was not available in 4-channel mode in 2025. Overall, we find that a weighted average of RV measurements across channels mitigates the effects of night-to-night shifts, recovering the Gaia-ESO catalog RV more consistently than any individual measurement alone. 

\subsection{Cram\'er-Rao Bound}

In the photon-limited regime, the maximal achievable RV precision that can be measured from a spectrum is given by an application of the Cram\'er-Rao bound \citep{cramer1946, rao1945} to a spectrum of independent Gaussian pixels \citep{bouchy_pepe_queloz_2001}:

\begin{equation}
    \sigma_v = \frac{c}{Q \sqrt{N_{e^-}}},
\end{equation}

\noindent where, given expected photoelectrons $A_0(i)$ and read noise $\sigma_D(i)$ in pixel $i$ at wavelength $\lambda_i$, the total number of photoelectrons $N_{e^-} \equiv \sum_i A_0(i)$ and the quality factor is defined as

\begin{equation}
    Q \equiv \frac{\sqrt{\sum_i W(i)}}{\sqrt{\sum_i A_0(i)}},\,\, W(i) \equiv \frac{\lambda_i^2 \left[\partial A_0(i) /\partial \lambda\right]^2}{A_0(i) + \sigma_D^2(i)}.
\end{equation}

To evaluate the theoretical photon-limited $\sigma_v$, we adopted a BOSZ spectrum representative of a typical G-type dwarf ($T_{\rm eff}=5500$ K, $\log g=4.5$, and $[\mathrm{M/H}]=0$), degraded it to the specified resolution, and sampled it on representative NGPS wavelength grids. We imposed the specified SNR by assigning SNR$^2$ expected photoelectrons to each continuum pixel, assuming Poisson noise and negligible read noise. The 25th, 50th, and 75th percentiles of the measured SNR distribution across our G-, R-, and I-channel spectra are approximately 32, 46, and 58, respectively. After masking telluric-dominated pixels, we found the Cram\'er--Rao bounds at $R=3000$ and SNR $= 30$, $45$, and $60$ to be approximately $(0.9,\,0.6,\,0.4)$ km s$^{-1}$ in the G channel, $(2.1,\,1.4,\,1.0)$ km s$^{-1}$ in the R channel, and $(2.1,\,1.4,\,1.1)$ km s$^{-1}$ in the I channel, respectively. At $R=4500$, these bounds improve to $(0.5,\,0.4,\,0.3)$, $(1.3,\,0.8,\,0.6)$, and $(1.5,\,1.0,\,0.7)$ km s$^{-1}$, respectively.

We conclude that decreasing slit width tends to improve the bound, but that the gain can be offset by decreased throughput. Furthermore, at SNRs typical of our observed spectra, the derived Cram\'er-Rao bounds are lower than the empirically recovered RV precisions, implying that the RV performance of NGPS is not photon-limited. Instead, the RV performance is primarily set by systematic uncertainties in calibration and modeling.

\section{Conclusion}
\label{sec:conclusion}

We have observed $24$ radial velocity (RV) standards from the Gaia-ESO survey with the Next Generation Palomar Spectrograph (NGPS) to calibrate the performance of the instrument as a stellar speedometer. We summarize our findings below.

\begin{itemize}
    \item NGPS experiences significant flexure, inducing shifts in the wavelength solution from epoch to epoch. Across the two observing nights, the median corrections applied to individual spectra were approximately $+35$ to $+42$ km s$^{-1}$ in the R channel and $-53$ to $-62$ km s$^{-1}$ in the I channel. If no corrections are applied, typical RVs are biased at the $\sim 50$ km s$^{-1}$ level, demonstrating that flexure calibration is essential for stable NGPS RV measurements.
    \item To correct for flexure, we develop a publicly available pipeline to measure (wavelength-dependent) flexure corrections in the G, R, or I channels. By default, we fit a multiplicative model incorporating both stellar and telluric template spectra to spectral regions predicted to have strong telluric absorption features. In the G channel, where telluric absorption is weak, we centroid the strong [O I] 5577\,\AA~sky line instead. We interpolate the resulting flexure measurements to derive robust RVs in spectral regions dominated by stellar features. We find that an inverse-variance-weighted average of the independently measured RVs from all available channels recovers the Gaia-ESO catalog RV to within $5$ km s$^{-1}$ in $\approx 97\%$ of observations, compared to $\approx 95\%$ of observations when using the I channel alone.
    \item Our pipeline retains the fallback option of measuring flexure based on sky emission lines in the R or I channels. However, we find that telluric absorption features perform considerably better than sky emission lines as a method of computing flexure corrections. The latter approach is less precise and accurate, since sky emission fills the slit and does not track the illuminating point source.
    \item We do not find a strong trend in RV performance with slit width when telluric absorption lines are used as a flexure source. Using a smaller slit width increases spectral resolution, but any gains can be offset by reduced throughput. We regard a $0.50''$ slit as the best general-purpose choice, with a smaller slit appropriate for bright sources, or when sky emission lines are used as a flexure source. We find that the RV performance of NGPS is limited by systematics in calibration and modeling, rather than photon statistics alone.
\end{itemize}

We have demonstrated that telluric absorption features can robustly correct for flexure while saving time relative to arc calibrations taken throughout the night. We recommend using the I channel to measure RVs, or taking an inverse-variance-weighted average of the independently measured RVs in each channel.

Future improvements could result from refining the NGPS wavelength solution or introducing a 2D noise model to the QuickLook Data Reduction Pipeline. In addition, the pipeline should be tested on a wider sample of targets with known RVs, including cool stars, hot stars, fast rotators, and luminous composite binaries. An active flexure compensation system is in development (Fremling et al.\ in prep.), though its performance has not yet been validated. We look forward to the important role that NGPS will play in spectroscopic follow-up campaigns of targets of interest discovered in current and upcoming large surveys, such as \textit{Gaia} DR4, Roman, or LSST.

\begin{acknowledgments}
This research was supported by NSF grants AST-2307232 and AST-2540180. This work has made use of data from the European Space Agency (ESA) mission {\it Gaia} (\url{https://www.cosmos.esa.int/gaia}), processed by the {\it Gaia}
Data Processing and Analysis Consortium (DPAC,
\url{https://www.cosmos.esa.int/web/gaia/dpac/consortium}). Funding for the DPAC has been provided by national institutions, in particular the institutions
participating in the {\it Gaia} Multilateral Agreement. We acknowledge the aid of ChatGPT 5.5 and 5.6 in code development.
\end{acknowledgments}

\begin{contribution}

PN was responsible for performing the observing and data analysis and writing the paper. KE came up with the initial research concept and obtained the funding. 


\end{contribution}

%
\facilities{Hale (NGPS)}

\software{astropy \citep{2013A&A...558A..33A, 2018AJ....156..123A, 2022ApJ...935..167A}, lacosmic \citep{lacosmic_zenodo}}


\clearpage
\appendix

\section{Spatial Variation of Sky Lines}
\label{sec:appendix}

\begin{figure*}[!ht]
    \centering
    \includegraphics[width=\textwidth]{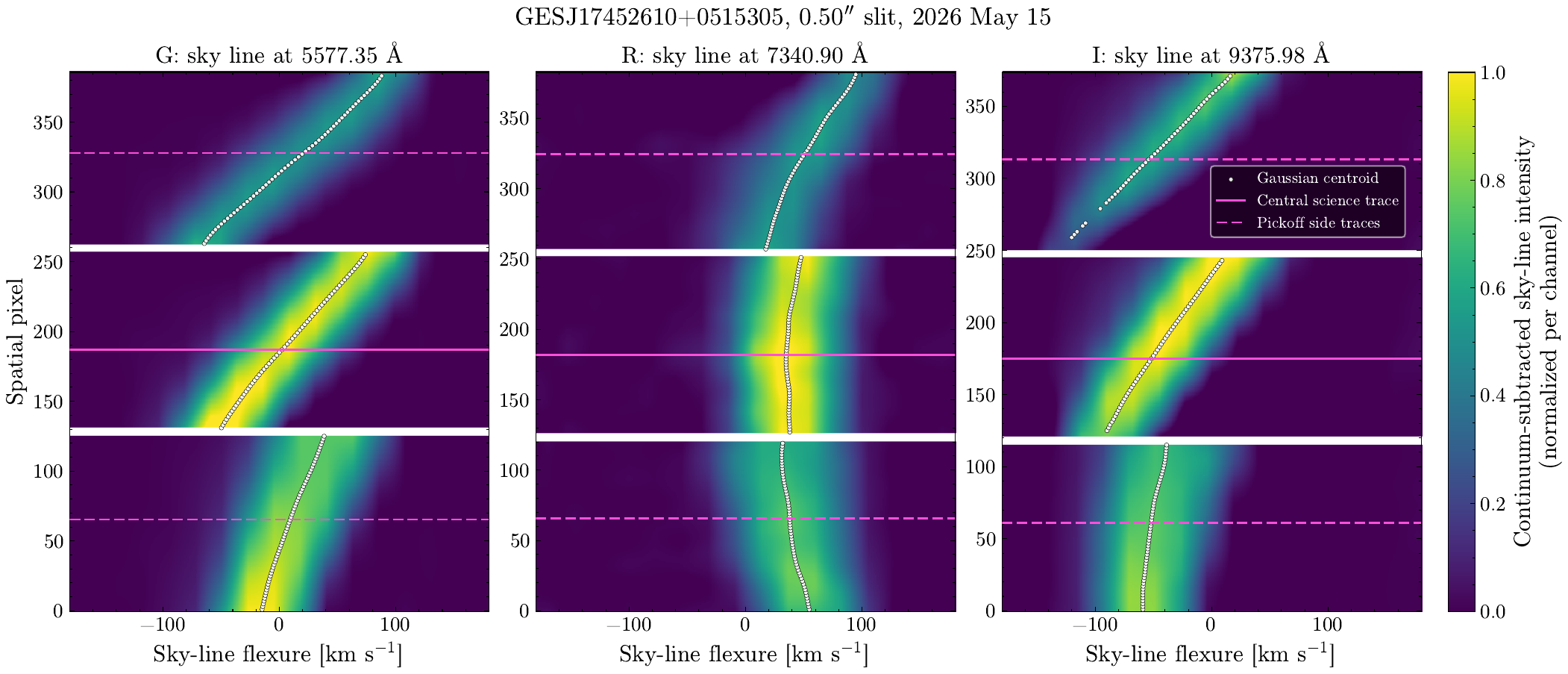}
    \caption{Spatial variation, or tilt, of representative strong sky emission lines in each channel. The heatmap indicates the intensity of each sky line. The solid magenta line shows the location of the science trace, while the dashed magenta lines indicate the location of the side traces. The white circles indicate the peak of the fitted Gaussian at each spatial position. The local spatial tilt at the location of the science trace is linear, but the fitted sky-line centroid varies non-linearly across the entire detector, with large jumps between image slices.}
    \label{fig:spatial_tilt}
\end{figure*}

We now investigate the flexure measured from sky emission lines as a function of spatial position on the NGPS detector. We plot the flexure measured from Gaussian fitting of representative strong sky lines in each channel in Figure~\ref{fig:spatial_tilt}, using a heatmap to visualize the sky line intensity in each of the image slices. We find that, close to the central pixel on the central slice (at the measured position of the science trace), the spatial tilt is locally linear. Fitting a straight line to the sky-line centroids over the five spatial rows used in the boxcar extraction, we measure local spatial gradients of approximately $+1.13$, $+0.065$, and $+0.849$ km s$^{-1}$ pixel$^{-1}$ in the G, R, and I channels, respectively. Summing together the rows in the boxcar extraction causes the spatial tilt to approximately cancel out. 

However, the spatial tilt varies significantly over the entire reduced 2D image; the fitted sky line centroids show non-linear variation across detector rows within each channel, with clear jumps between slices. Since the traces in the side slices are significantly fainter than the central science trace (i.e., they originate from light surrounding the central slit that is redirected by ``pickoff'' mirrors), they contribute less useful information towards measuring the stellar RV. Furthermore, co-addition of the slices is complicated by slice-dependent line-spread functions and small relative shifts in their wavelength calibrations. Thus, we defer a joint forward model of the central science trace together with the fainter side traces to future work.

\clearpage


\bibliography{bibliography}{}
\bibliographystyle{aasjournalv7_modified}



\end{document}